\documentclass[prl,amsfonts,amssymb,floats,twocolumn,aps]{revtex4-2}

\usepackage[final,dvips]{epsfig}
\usepackage{amsmath,amsfonts}
\usepackage{bbm}
\usepackage{float}
\usepackage[caption = false]{subfig}
\usepackage{graphicx}
\usepackage{color}
\usepackage{hyperref}
\usepackage[normalem]{ulem}

\newcommand{\kk}{\vec{k}}
\newcommand{\kp}{\vec{K}}
\newcommand{\kpp}{\vec{K}'}
\newcommand{\slr}{\sigma_{xx}^{R}}
\newcommand{\sli}{\sigma_{xx}^{I}}
\newcommand{\shr}{\sigma_{xy}^{R}}
\newcommand{\shi}{\sigma_{xy}^{I}}

\usepackage{pdfpages}
\usepackage{pgffor}
\makeatletter
\AtBeginDocument{\let\LS@rot\@undefined}
\makeatother

\begin{document}
\title{Photonic spin-Hall effect as a probe for time-reversal-symmetry broken  band topological phases}

\date{\today}

\author{Deblina Samanta}
\affiliation{Tata Institute of Fundamental Research, Hyderabad 500046, India}

\author{Darshan G. Joshi}
\affiliation{Tata Institute of Fundamental Research, Hyderabad 500046, India}

\begin{abstract}
When a plane polarized Gaussian beam of light is incident on a surface, upon reflection it splits into right and left circularly polarized beams that are spatially separated in the direction perpendicular to the plane of incidence. This is known as the photonic spin-Hall effect (PSHE).  
In this work, we show that the centroid shift, which is the intensity weighted average of the shifts of the right and the left circularly polarized beams directly probes the optical Hall conductivity, which carries the essential information about the topological properties of the system. We show that the centroid shift as a function of the frequency of light has a unique sign structure depending on whether the system is in a time-reversal symmetry broken band topological phase or a trivial phase. 
Thus, the PSHE may serve as an unambiguous and a non-invasive probe to detect time-reversal symmetry broken band topological phases. 
\end{abstract}

\maketitle 

{\em Introduction.--} Topology of electronic bands has been a subject of continued interest in modern condensed-matter physics \cite{Hasan_Kane, Qi_Zhang}. These ideas have lead to the discovery of a plethora of band topological phases which are beyond the standard Landau's paradigm. Topological phases are found all the way from a simple $2d$ electron gas in a magnetic field \cite{QHE}, to cold atoms \cite{Jotzu2014}, and more recently in the Moir\'{e} materials \cite{Zhang_Senthil}. In $2d$ these band topological phases are best understood in the presence of a bulk band gap, such as in the case of a Chern insulator akin to quantum-Hall effect, topological insulators and topological superconductors \cite{Hasan_Kane, Qi_Zhang}. One of the hallmarks of these topological phases is a bulk topological invariant leading to non-trivial edge states via bulk-boundary correspondence. An observable consequence is a quantized Hall conductivity in the case of Chern insulators \cite{TKNN}. In the case of quantum spin-Hall effect or topological insulator, the Hall conductivity vanishes due to time-reversal symmetry, but the spin-Hall conductivity is quantized \cite{Kane_Mele}. Similar transport measurement related predictions exists for different topological phases, both in $2d$ as well as $3d$. It is also important to note that such quantized responses wherever possible are usually in the DC limit. However, the quantized magneto-optical response is an example in which the quantization is present for a range of low frequencies \cite{Tse_MacDonald}.

Although electrical transport measurements are one of the standard and often insightful measurements, there may be several practical challenges. 
It is therefore desirable to have non-invasive probes to detect topological phases. 
In this regard, an optical probe is a viable alternative \cite{Moe_review}. As long as the power of the incident light is not very high it can serve as a non-invasive probe. The magneto-optical Kerr effect is one of such probes which is routinely used to characterize magnetic thin films \cite{Gong2017, Sivadas2016}. It has also been discussed in the context of topological insulators with weak time-reversal symmetry breaking \cite{Tse_MacDonald, Tse2011, Aguilar2012}. In certain hexagonal semiconductors, light absorption has  been shown to be valley-dependent polarization sensitive resulting in selection rules that may lead to selective photoexcitation of carriers \cite{Xiao2012}.  

Most of the optical probes rely on the gapless (or weakly gapped) surface states in a thin layer of topological insulator. Thus the response is typically from the non-trivial surface originating from underlying topology. While this is of great advantage, it may not be particularly suitable in cases where there are no surface Dirac cones. For instance, $2d$ Chern insulator has only gapless edge modes which would not participate in typical light probes. Also, Weyl semimetals have only surface Fermi arcs. Therefore, it may be advantageous to have an optical probe that depends on the bulk properties in order to detect topology of electronic bands. At first instance this may sound counter-intuitive since one of the hallmarks of band topology are the surface or edge modes.
In this work, we show that the photonic spin-Hall effect of light is a sensitive probe of the optical Hall conductivity, which encodes the essential information about the band topology. This bulk probe at available optical frequency can distinguish between a trivial and topological phase in the case of time-reversal symmetry breaking.

\begin{figure}[t]
\centering
\includegraphics[width=0.45\textwidth]{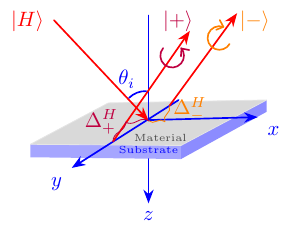}
\caption{Schematic of the photonic spin-Hall effect where an incident Gaussian wavepacket of light polarized in the $H$ plane (here $x-z$ plane) upon reflection splits into two beams of opposite circular polarizations that are spatially separated along the $y$ axis, denoted by $\Delta^{H}_{\pm}$.
}
\label{fig:pshe}
\end{figure}

{\em PSHE.--} The photonic spin-Hall effect (PSHE) \cite{Onoda_2004, Ling_2017}, also known as  Imbert-Fedorov shift \cite{Bliokh_2007, Bliokh_2013}, is a phenomena where a plane-polarized wavepacket (typically a Gaussian beam) of light upon reflection from a surface {\em spatially splits} into two beams of opposite circular polarizations (see Fig. \ref{fig:pshe}) in the direction perpendicular to the plane of incidence. Note that there is also a similar shift in the direction parallel to the plane of incidence, known as Goos-H\"{a}nchen shift \cite{Bliokh_2013}, which we shall not consider in this work. This splitting or the shift is typically of the order of the wavelength of the incident light. However, using the technique of weak measurement amplification \cite{Bernardo2014, Ling_2017} it has been recently possible to measure these shifts accurately \cite{Panda2022}. The PSH shift is a consequence of the Berry curvature arising in the momentum space of the light \cite{Onoda_2004, Bliokh_2013}. Since the wavepacket is composed of multiple momenta and thus polarization vectors, a non-trivial Berry curvature arises as the polarization vector rotates in a closed loop in the momentum space. It has been shown that this shift is a function of the Fresnel coefficients, which themselves depend on the optical conductivity of the material \cite{Onoda_2004, Bliokh_2013}. The optical conductivity carries a fingerprint of the quantum phase of the material. Thus, a natural question is whether the PSH shift could be used as a probe for quantum phase of the material. In this work, we demonstrate that indeed PSH shift can unambiguously distinguish between a topological and a trivial quantum phase. We explicitly demonstrate this for two important cases: Chern insulator in $2d$ and Weyl semimetal in $3d$.

The PSHE has been calculated for semiconductors \cite{Ling_2017, Da2023} and in particular for Graphene in external magnetic field \cite{Kort_Kamp} it has been predicted to take quantized values. Apart from semiconductors, PSH shift has also been predicted for Weyl semimetals \cite{Da2021} as well as for metamaterials and nanoparticles in optical traps \cite{Roy_2014}. Experimentally, the PSH shift has been measured in the case of semiconductors \cite{Panda2022}, metamaterials \cite{Yin2013}, and also used to make precision measurement of optical conductivity \cite{Chen2020}.  Recently, it has been also calculated in the Haldane model \cite{PhysRevB.109.235418} and across topological phase transition \cite{Kort-Kamp_PRL} using low-frequency analysis. However, a systematic study of the PSHE to probe the quantum phase has been lacking. Importantly, it has been so far unclear whether PSHE could be used as a probe to detect a particular quantum phase unambiguously. 
In addition, the full dependence of the PSH shift on optical conductivity has not been explicitly made clear analytically. So far the identification has been mainly either by comparing specific features like peaks in the optical conductivity with those in the PSH shift or by certain low-frequency analysis \cite{Kort-Kamp_PRL}. 
Such analyses has not revealed the role of optical Hall conductivity in the PSHE.
In this work, we shall show explicitly how the optical Hall conductivity plays a crucial role in the PSHE and how it can be directly used to probe time-reversal symmetry broken topological phases. 

The PSH shift for the reflected left and right circular polarization with incident plane polarized beam in the $H$ plane (i.e. in the plane of incident light, which is $x-z$ plane here) in terms of the Fresnel reflection coefficients is \cite{Chen_2018, Liu:20}
\begin{equation}
\Delta_{\pm}^H = \mp\frac{\cot{\theta_i}}{k_0} ~Re\Bigg[\frac{r_{pp}+r_{ss}}{r_{pp}\mp  i r_{sp}}\Bigg] \,,
\label{eq:sh}
\end{equation}
where $\theta_{i} \in (0,\pi/2)$ is the incident angle, $k_{0}$ is the central wavenumber of the incident Gaussian beam, and $r's$ are the Fresnel reflection coefficients whose expressions can be found in the Suppl. Matt. \cite{SM}. The subscripts $p$ and $s$ in the Fresnel coefficients refer to the polarizations in the plane parallel and perpendicular to the incident plane respectively. The expression for $\Delta_{\pm}^{H}$ in terms of the optical conductivities can be also found in the Suppl. Matt. \cite{SM}. Analogously, one can also define the shift when the polarization of the incident light is in the plane perpendicular to the incident plane (denoted as $V$ plane). In this work we only consider the incident polarization in the $H$ plane, but our results also apply to the case when the incident light is polarized in the $V$ plane.

It turns out that the PSH shift of the RCP and LCP light, $\Delta_{\pm}^{H}$, does not provide a direct probe of the Hall conductivity and hence that of topology (see the detailed expressions in Eqs. (S36)-(S38) in the Supp. Matt. \cite{SM}).
We instead consider the intensity weighted average of $\Delta_{\pm}^{H}$, called the centroid shift,
\begin{equation}
\Delta_{c} = \frac{I_{+} \Delta_{+}^{H} + I_{-} \Delta_{-}^{H}}{I_{+} + I_{-}}    \,,
\label{eq:csi}
\end{equation}
where $I_{\pm}$ are the intensities of the reflected RCP and LCP beams. In terms of the Fresnel coefficients, the expression for the centroid shift is given by 
\begin{equation}
\Delta_{c} = -\frac{\cot{\theta_i}}{k_0} ~Re\Bigg[\frac{i (r_{pp}+r_{ss})r_{sp}^*}{|r_{pp}|^2+|r_{sp}|^2} \Bigg] \,,
\label{eq:cs}
\end{equation}
and the corresponding expression in terms of the optical conductivities is
\begin{equation}
\Delta_{c} =-
\frac{g_{1} \cot{\theta_i} \left[g_2 (\slr \shi -\sli\shr) + g_3 \shi \right]}{k_{0} D_{c}} \,,
\label{eq:cs_cond}    
\end{equation}
where $g's$ are $\theta_{i}$-dependent positive constants and $D_{c} \equiv D_{c}(\theta_{i},\omega) >0$ (see Suppl. Matt. \cite{SM} for expressions). This is the central result of our work.  
It is this quantity on which we will focus our attention and show that its frequency dependence has a specific sign structure depending on the topological phase. Note that the centroid shift ($\Delta_{c}$) is directly proportional to the Hall conductivity and so in the presence of time-reversal symmetry it is trivially zero. This is not the case for $\Delta_{\pm}$. In fact, so far the role of Hall conductivity in $\Delta_{\pm}$ has been unclear.
In this work, we have revealed explicitly the role of optical Hall conductivity in PSH shift, and shown that in fact the centroid shift as defined above is sensitive to it. At the Brewsters angle $(\theta_{B})$, the constants $g'$s in Eq. (\ref{eq:cs_cond}) take a particularly simple form (see Supp. Matt. \cite{SM}). Therefore, we will present explicit results at the Brewsters angle as this is also experimentally relevant. However, our results are general and valid at all incident angles. In the Supp. Matt. \cite{SM} we also present our results for different incident angles and the main conclusions are independent of the incident angle. 

We shall now demonstrate our results in three important classes of band topological phases: (i) Chern insulator in $2d$, (ii) Weyl semimetal in $3d$ and (iii) Quantum anomalous Hall insulator in $3d$.

\begin{figure*}[t]
\centering
\subfloat[]{\includegraphics[width=0.30\textwidth]{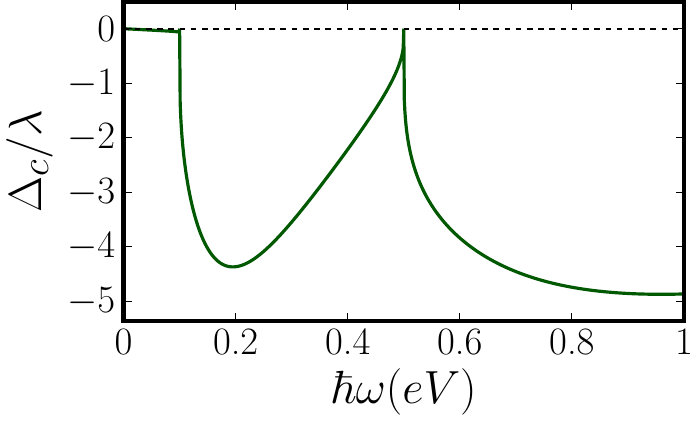}}
\subfloat[]{\includegraphics[width=0.29\textwidth]{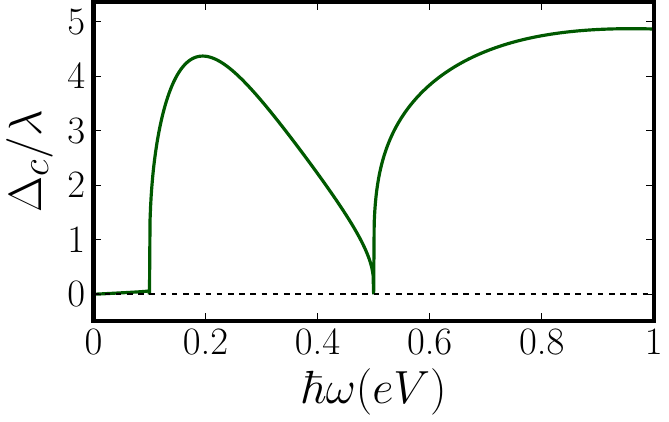}} 
\subfloat[]{\includegraphics[width=0.30\textwidth]{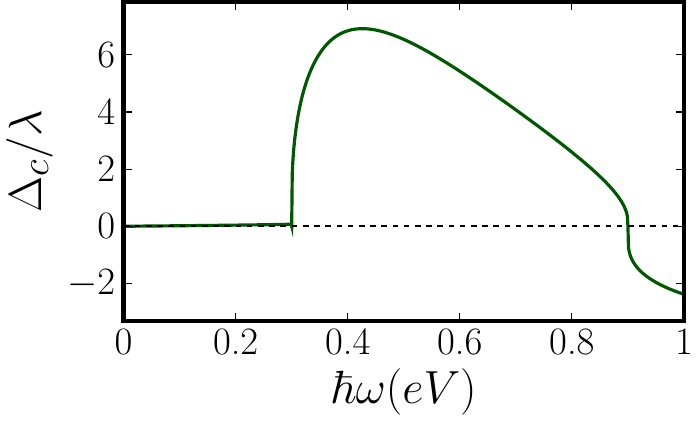}} 
\caption{Centroid shift in the Chern insulator and trivial phases. (a) $M=0.1 eV\,, ~ \phi=\pi/2$ with $C=1$. (b) $M=0.1 eV\,, ~ \phi=-\pi/2$ with $C=-1$. (c) $M=0.3 eV\,, ~ \phi=\pi/2$ with $C=0$.  
}
\label{fig:cs}
\end{figure*}

{\em Chern insulator.--} We consider the Haldane model \cite{Haldane} on a honeycomb lattice, which is a paradigmatic model for Chern insulator. The Hamiltonian reads, 
\begin{equation}
H = -t\sum_{\langle ij\rangle} c^{\dagger}_{i} c_{j} + t' \sum_{\langle\langle ij\rangle\rangle} e^{i\phi_{ij}} c^{\dagger}_{i} c_{j} + \sum_{i} \mu_{i} c^{\dagger}_{i} c_{i} \,,
\label{eq:ham}
\end{equation}
where $t$ is the nearest-neighbor hopping amplitude, $t'$ is the next-nearest-neighbor hopping amplitude, $\phi_{ij}$ is the phase which is positive for clockwise hopping to second neighbor, $\mu_{i}$ is the site-dependent chemical potential. In the absence of the second-neighbor hopping and uniform chemical potential, this model is simply the tight-binding model of fermions on a honeycomb lattice displaying Dirac points at the corners of the Brillouin zone ($\vec{K}$ and $\vec{K}'$ points). With the breaking of inversion symmetry by the introduction of a staggered potential $\mu_{A} \neq \mu_{B}$ on the two sublattices a gap opens. On the other hand, breaking of the time-reversal symmetry by introducing the second-neighbor complex hopping ($t'$ term in Eq. \ref{eq:ham} with $\phi \neq 0$) also opens a gap. These two mechanisms of opening the gap have fundamentally different effect: Only inversion symmetry breaking leads to a trivial insulator, whereas time-reversal symmetry breaking leads to a Chern insulator, characterized by a quantized Hall conductivity, in an extended region of parameter space. The relevant low-energy physics is captured by the Hamiltonian in the vicinity of the $\kp$ and $\kpp$ points,
\begin{equation}
h_{\kk} = -\lambda \mathbbm{1} + \hbar v_f (\tau k_x \sigma_{x} + k_y \sigma_{y}) + \Delta_\tau \sigma_{z} \,, 
\label{eq:lh}
\end{equation}
where $\tau=\pm 1$ corresponds to the $\kp$ and $\kpp$ points or valley respectively and $\kk$ is the momentum away from these points. Further, $v_f=\sqrt{3} ta/2 \hbar$, with $a$ being the lattice constant, $\lambda=-(\mu_{A} + \mu_{B})/2 +3 t'\cos{\phi} $, $\Delta_\tau= M - \tau 3\sqrt{3} t' \sin{\phi}$, with $M = (\mu_{A} - \mu_{B})/2$. This model has a trivial phase when $sgn(\Delta_{\pm 1})$ is the same, while it is in the Chern insulator phase when $sgn(\Delta_{1}) \neq sgn(\Delta_{-1})$.

The trivial and the topological Chern insulator phases can be distinguished by the DC Hall conductivity, $\sigma_{xy}(\omega=0) = C e^{2}/h$, where the Chern number $C=0$ in the trivial phase and $C= \pm 1$ in the topological phases. The optical conductivity for the Haldane model can be worked out using the low-energy Hamiltonian \cite{PRB101}, and detailed expressions can be found in the Suppl. Mat. \cite{SM}.

Let us now look at the centroid shift (Eq. (\ref{eq:cs_cond})) for the Haldane model. We will focus on the term in square brackets in the numerator of Eq. (\ref{eq:cs_cond}) because all other terms are always positive. 
Firstly, we note that $\slr(\omega) \ge 0 $ and $\sli (\omega) \leq 0$ for all values of $\omega$ in both the topological and the trivial phases (see Supp. Matt. \cite{SM} for expressions and plots). Secondly, in the $C=1$ phase $\shr, \shi \ge 0$, while in the $C=-1$ phase $\shr, \shi \le 0$. Thus $\Delta_{c} \le 0$ for $C=1$ and $\Delta_{c} \ge 0$ for $C=-1$ for all $\omega$. However, in the trivial phase, $\shr$ and $\shi$ have opposite signs in the region, $ \min{(|\Delta_{1}|, |\Delta_{-1}|)} \leq \hbar\omega/2 \leq \max{(|\Delta_{1}|, |\Delta_{-1}|)}$. This also follows from the optical sum rules, which relate the integral of the Hall conductivity to the Chern number.   Therefore, $\Delta_{c}$ necessarily changes sign in this region in the trivial phase. In Fig. (\ref{fig:cs}) we show the centroid shift in the trivial and the topological Chern insulator phases.

So the main result is that the centroid shift as a function of frequency has the same sign in the topological Chern insulator phase, whereas it changes sign as a function of frequency in the trivial insulator phase. This peculiar sign structure thus  provides a conclusive probe for the presence or absence of a Chern insulator phase. 

\begin{figure*}[t]
\centering
\subfloat[]{\includegraphics[width=0.33\textwidth]{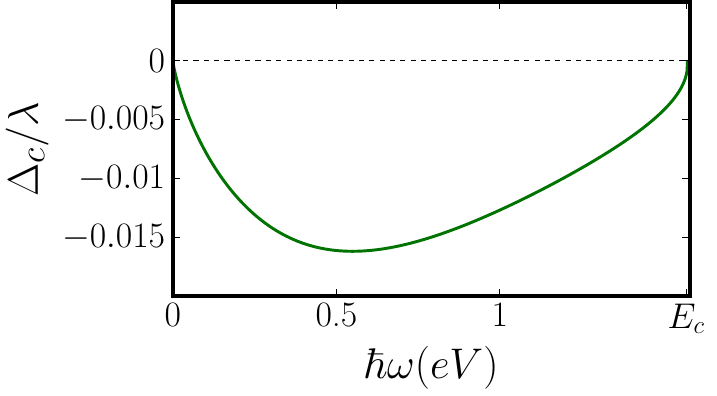}}
\subfloat[]{\includegraphics[width=0.33\textwidth]{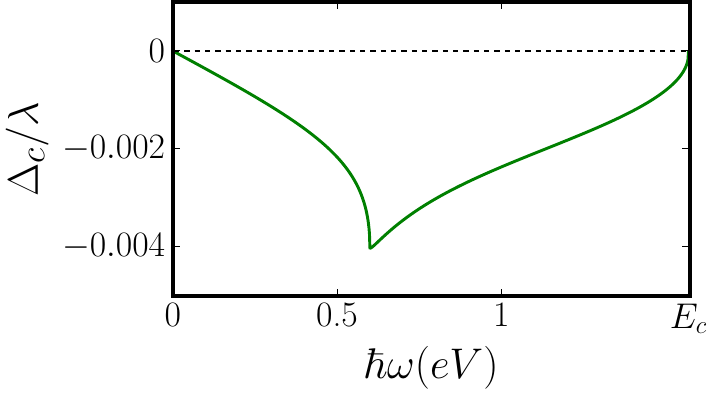}}
\subfloat[]{\includegraphics[width=0.33\textwidth]{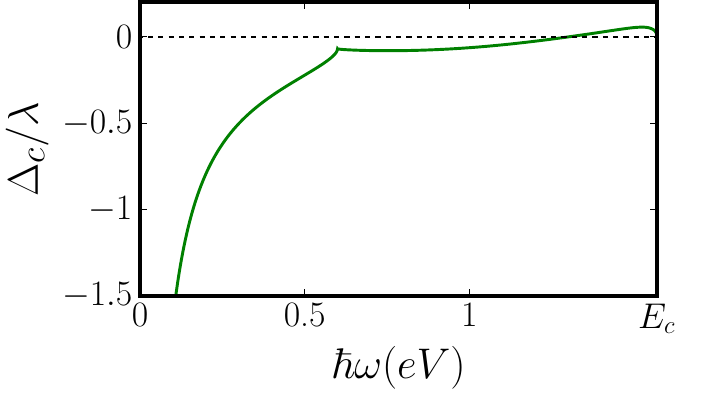}} 
\caption{(a) Centroid shift in the WSM phase with the incident plane being the $x-y$ plane. (b) Centroid Shift in $x-y$ plane for QAH phase. (c) Centroid Shift in $x-y$ plane for NI phase. Centroid shifts in the  $x-z$ and $y-z$ planes are zero for all the phases.
For plots (b) and (c), we have set $|\alpha| = 0.3$ eV, $|\beta| = 0.5 a^{2}/2$ eV$\AA^2$ with $a=2\AA$. Here $E_{c}$ is the high-energy cut-off.}
\label{fig:cs3d}
\end{figure*}

{\em Weyl semimetal.--} 
Next we consider the Weyl semi-metal phase in $3d$ with broken time-reversal symmetry. The effective low-energy Hamiltonian for the Weyl semimetal (WSM) phase around one of the Weyl nodes is as follows \cite{PhysRevB.95.161112}\cite{Chen_2018}\cite{Liu:20}:
\begin{equation}
\label{eq:hwsm}
H = \varepsilon \left(k_{-} \sigma_{+} + k_{+} \sigma_{-} \right)
+ \hbar v_z (k_z \mp b)\, \sigma_z \,,
\end{equation}
where $k_{\pm} = k_{x} \pm i k_{y}$, $\sigma_{\pm} = (\sigma_{x} \pm i \sigma_{y})/2$, with $\sigma$'s being the Pauli matrices, while $\varepsilon$ and $v_{z}$ are material dependent parameters. 
Without loss of generality, we have considered the Weyl nodes along the $k_{z}$ axis, separated by $b$ and at the same zero energy.

Since the Weyl nodes are along the $k_{z}$ axis the Hall conductivity is non-zero only in the $x-y$ plane. Consequently, the centroid shift, $\Delta_{c}$ (Eq. \ref{eq:cs_cond}), is identically zero for $x-z$ and $y-z$ planes. 
Moreover, in the $x-y$ plane, the centroid shift has the same sign as a function of frequency (see Fig. \ref{fig:cs3d}(a)), just as in the case of Chern insulator phase in $2d$. This can be straightforwardly obtained using the optical conductivity expressions in WSM phase, whose details can be found in the Supp. Mat. \cite{SM}. These results have two implications. Firstly, using PSHE one can determine the direction along which the Weyl nodes are located since only the plane normal to the axis passing through the Weyl nodes will lead to a non-vanishing $\Delta_{c}$. Secondly, it clearly distinguishes the WSM phase (with broken time-reversal symmetry) from another possible gapless phase, namely $3d$ Dirac semimetal. In the Dirac SM phase the time-reversal symmetry is preserved and so the Hall conductivity identically vanishes on all the planes leading to vanishing centroid shift. 

{\em Quantum anomalous Hall phase.--} Now we turn our attention to gapped phases in $3d$. To demonstrate the utility of the PSHE, we consider the quantum anomalous Hall (QAH) phase and the normal insulator (NI) phase. These two phases can be studied using the following effective low-energy Hamiltonian \cite{PhysRevB.95.161112}:
\begin{equation}
\label{eq:hqah}
H = \varepsilon \left(k_{-} \sigma_{+} + k_{+} \sigma_{-} \right)
+ (\alpha+\beta q_z^2)\, \sigma_z \,,
\end{equation}
where $\alpha,\beta < 0$ in the QAH phase, while $\alpha,\beta > 0$ in NI phase. Additionally, $q_z = k_z$ and $q_z = k_z \mp \pi/a$ in the NI and QAH phases respectively. The details of the optical conductivities in both these phases is given in the Supp. Mat. \cite{SM}. Using these results we find that again in the topological phase, i.e. QAH phase, 
the centroid shift doesn't change sign as a function of frequency (see Fig. \ref{fig:cs3d} (b)). On the other hand, in the trivial insulator phase, i.e. NI phase, there is a sign change in $\Delta_{c}$.
Thus PSHE can distinguish unambiguously between these gaped phases in $3d$. 

{\em Discussion.--} 
In this work, we show that the centroid shift using PSHE can unambiguously identify time-reversal-symmetry broken band topological phases. 
In the topological phase, the frequency dependence of the centroid shift has the same sign, while in the trivial phase there is a definite sign change. This is guaranteed by the optical sum rules. 
Typically, the information of Chern number is extracted via dc measurements only. Here we see that optical measurements can be used as a direct probe of a non-zero Chern number. This technique and protocol may be  particularly useful where transport measurements may not be able to provide conclusive evidence, such as in the Weyl semi-metals. Importantly, the PSHE is a non-invasive probe. So from an experimental point of view it can be used in cases where sample geometry and size may be a limitation for transport measurements. 

A straightforward experimental protocol can be devised based on our results. It is not necessary to scan the full frequency range. In the case of $2d$, one measurement at a  low frequency ($\hbar\omega < 2\min{(|\Delta_{1}|, |\Delta_{-1}|)}$) and another at a higher frequency ($\hbar\omega > 2\max{(|\Delta_{1}|, |\Delta_{-1}|)}$) is sufficient. If the sign of $\Delta_{c}$ at both these frequencies is the same then it is a Chern insulator phase; otherwise it is a trivial phase. Similarly, in  $3d$, it is sufficient to measure PSHE at an energy less than the bulk gap and at sufficiently high energy (but less than cutoff energy $E_{c}$). Again, the sign structure of the centroid shift can unambiguously distinguish between the trivial insulator and topological quantum anomalous Hall phases. For gapless WSM phase, PSHE can additionally also clearly determine the axis along which the Weyl nodes are located.

Although in this work we have explicitly demonstrated our results in the case of non-interacting band topological phases, our strategy is quite general. We envision the application of these results even in the case of time-reversal-symmetry broken interacting systems such as fractional Chern insulators. Our work opens wide possibilities of future directions using this non-invasive probe for topological phases. Moreover, the centroid shift using the PSHE need not be restricted to probing just topology.
Our strategy could be extended to phases where time-reversal is spontaneously broken, especially magnetic phases. Whether PSHE could be used to further reveal non-trivial magnetic textures is an interesting direction which will be pursued in future. 

An immediate extension of our work is possible in the case of multi-band systems typically arising in many semiconductors as well as magnetically doped semiconductors. This is very relevant for the spintronics and valleytronics applications.

Since the PSHE can be measured quite accurately in different settings \cite{Yin2013, Chen2020, Panda2022}, our results could pave a way to utilize it as an important measurement to characterize  different quantum phases.

{\em Acknowledgment.--} We thank G. Rajalakshmi, J. J. Panda, T. N. Narayanan and Lakshmi N. Govind for helpful discussions. 
We acknowledge the support of the Department of Atomic Energy, Government of India, under Project Identification No. RTI 40007.

\bibliography{pshe_chern}


\newpage


\section*{Supplementary material}

# Supplemental material:
# Photonic spin-Hall effect as a probe for time-reversal-symmetry broken band topological phases

Deblina Samanta and Darshan G. Joshi
*Tata Institute of Fundamental Research, Hyderabad 500046, India*

## I. FRESNEL COEFFICIENTS

The Fresnel reflection coefficients are obtained using the boundary conditions on electric and magnetic fields at the reflection surface. Importantly, these Fresnel coefficients depend on the optical conductivity of the reflecting material. The frequency dependence of the optical conductivity is dictated by the quantum phase of the material. Thus, the Fresnel coefficients inherit the information of the quantum phase of the material, in particular the topological property. Below we briefly quote the results for the Fresnel coefficients in terms of the optical conductivity (detailed derivations can be found in Refs. [S1–S3]):

$$r_{ss} = -\frac{a_-^s a_+^p + b^2}{a_+^s a_+^p + b^2}\,, \tag{S1}$$

$$r_{pp} = \frac{a_+^s a_-^p + b^2}{a_+^s a_+^p + b^2}\,, \tag{S2}$$

$$r_{sp} = -2\frac{Z_0^2}{Z_1}\frac{\mu_1\mu_2}{\mu_0^2}\frac{k_{z,1}k_{z,2}\sigma_{sp}}{a_+^s a_+^p + b^2}\,, \tag{S3}$$

$$r_{ps} = 2\frac{Z_0^2}{Z_1}\frac{\mu_1\mu_2}{\mu_0^2}\frac{k_{z,1}k_{z,2}\sigma_{ps}}{a_+^s a_+^p + b^2}\,, \tag{S4}$$

where,

$$a_\pm^p = \frac{k_{z,1}\epsilon_2 \pm k_{z,2}\epsilon_1 + k_{z,1}k_{z,2}\frac{\sigma_{pp}}{\omega}}{\epsilon_0}\,, \tag{S5}$$

$$a_\pm^s = \frac{k_{z,2}\mu_1 \pm k_{z,1}\mu_2 + \omega\mu_1\mu_2\sigma_{ss}}{\mu_0}\,, \tag{S6}$$

$$b^2 = -\frac{Z_0^2\mu_1\mu_2 k_{z,1}k_{z,2}\sigma_{sp}\sigma_{ps}}{\mu_0^2}\,. \tag{S7}$$

In the above expressions, the index $p$ denotes the direction parallel to the projection of the incident light propagation wavevector in the $2d$ material plane and $s$ denotes the direction perpendicular to it. We consider the reflection of light from a $2d$ $x-y$ plane at $z=0$. So, for conductivity, $p$ denotes the direction parallel to the component of the light propagation vector in the $x-y$ plane and $s$ the direction perpendicular to it. For brevity we have suppressed the frequency argument of the optical conductivity ($\sigma_{ij} \equiv \sigma_{ij}(\omega)$). The optical conductivities appearing in the above expressions are related to the standard definitions in the cartesian coordinates via a simple coordinate transformation as follows:

$$\begin{aligned}
\sigma_{pp} &= \sigma_{xx}\cos^2\phi + \sigma_{yy}\sin^2\phi + (\sigma_{xy}+\sigma_{yx})\sin\phi\cos\phi\,,\\
\sigma_{ss} &= \sigma_{xx}\sin^2\phi + \sigma_{yy}\cos^2\phi - (\sigma_{xy}+\sigma_{yx})\sin\phi\cos\phi\,,\\
\sigma_{ps} &= (\sigma_{xx}-\sigma_{yy})\sin\phi\cos\phi - (\sigma_{xy}\cos^2\phi - \sigma_{yx}\sin^2\phi)\,,\\
\sigma_{sp} &= (\sigma_{xx}-\sigma_{yy})\sin\phi\cos\phi + (\sigma_{xy}\sin^2\phi - \sigma_{yx}\cos^2\phi)\,.
\end{aligned} \tag{S8}$$

In the above relations, $\phi$ is the angle between the incident plane and the $x-z$ plane. Specifically, we consider only isotropic systems in this work and set the angle $\phi=0$. In isotropic materials, $\sigma_{xx}=\sigma_{yy}$ and $\sigma_{xy}=-\sigma_{yx}$. Also, $Z_n=\sqrt{\mu_n/\epsilon_n}$, and $k_{z,n}=\omega\sqrt{\epsilon_n\mu_n}\cos\theta_n$, with $\epsilon_n$ being the permittivity and $\mu_n$ being the permeability, and where the index $n=0,1,2$ denotes the free space, incident medium and the transmitted medium respectively. In our calculation we have considered the incident medium to be vacuum ($\epsilon_1=\epsilon_0$ and $\mu_1=\mu_0$) and the transmitted medium to be a non magnetic material (so $\mu_2=\mu_0$). Using the Snell's law we can express the Fresnel coefficients in terms of the incident angle ($\theta_i$) and the optical conductivities for the isotropic case as follows:

$$r_{pp} = \frac{(A_1+B_1\sigma_{xx})(C_1+D_1\sigma_{xx})+E_1\sigma_{xy}^2}{(A_1+B_1\sigma_{xx})(C_2+D_1\sigma_{xx})+E_1\sigma_{xy}^2}\,, \tag{S9}$$

$$r_{ss} = -\frac{(A_2+B_1\sigma_{xx})(C_2+D_1\sigma_{xx})+E_1\sigma_{xy}^2}{(A_1+B_1\sigma_{xx})(C_2+D_1\sigma_{xx})+E_1\sigma_{xy}^2}\,, \tag{S10}$$

$$\begin{aligned}
r_{sp} &= -\frac{E_2\sigma_{xy}}{(A_1+B_1\sigma_{xx})(C_2+D_1\sigma_{xx})+E_1\sigma_{xy}^2}\\
&= r_{ps}\,,
\end{aligned} \tag{S11}$$

where

$$A_1 = \sqrt{n_2^2-\sin^2\theta_i}+\cos\theta_i\,, \tag{S12}$$

$$A_2 = \sqrt{n_2^2-\sin^2\theta_i}-\cos\theta_i\,, \tag{S13}$$

$$B_1 = \mu_0 c\,, \tag{S14}$$

$$C_1 = \epsilon_0 c\left(n_2\cos\theta_i - \sqrt{1-\frac{\sin^2\theta_i}{n_2^2}}\right)\,, \tag{S15}$$

$$C_2 = \epsilon_0 c\left(n_2\cos\theta_i + \sqrt{1-\frac{\sin^2\theta_i}{n_2^2}}\right)\,, \tag{S16}$$

$$D_1 = \cos\theta_i\sqrt{1-\frac{\sin^2\theta_i}{n_2^2}}\,, \tag{S17}$$

$$E_1 = \mu_0 c\cos\theta_i\sqrt{1-\frac{\sin^2\theta_i}{n_2^2}}\,, \tag{S18}$$

$$E_2 = 2\cos\theta_i \sqrt{1 - \frac{\sin^2\theta_i}{n_2^2}}\,. \tag{S19}$$

In particular, at the Brewster angle ($\theta_B = \arctan n_2$),

$$A_1\big|_{\theta_i=\theta_B} = \sqrt{n_2^2+1}\,, \tag{S20}$$

$$A_2\big|_{\theta_i=\theta_B} = \frac{n_2^2-1}{\sqrt{n_2^2+1}}\,, \tag{S21}$$

$$B_1\big|_{\theta_i=\theta_B} = \mu_0 c\,, \tag{S22}$$

$$C_1\big|_{\theta_i=\theta_B} = 0\,, \tag{S23}$$

$$C_2\big|_{\theta_i=\theta_B} = \frac{2n_2\epsilon_0 c}{\sqrt{n_2^2+1}}\,, \tag{S24}$$

$$D_1\big|_{\theta_i=\theta_B} = \frac{n_2}{n_2^2+1}\,, \tag{S25}$$

$$E_1\big|_{\theta_i=\theta_B} = \frac{\mu_0 c n_2}{n_2^2+1}\,, \tag{S26}$$

$$E_2\big|_{\theta_i=\theta_B} = \frac{2n_2}{n_2^2+1}\,. \tag{S27}$$

The Fresnel reflection coefficients have some generic properties. At resonance frequencies corresponding to the energy band gap, $r_{pp} = 1, r_{ss} = -1, r_{sp} = r_{ps} = 0$.

## II. PHOTONIC SPIN HALL SHIFT

We consider a plane polarized Gaussian beam of light incident on a surface in the $x-y$ plane (as shown in Fig. 1 of the main text) located at $z = 0$. The electric field for a Gaussian beam is given by [S4, S5]

$$\vec{E}_s(\vec{r}) = \int dk_{sx}dk_{sy}\ \vec{\mathcal{E}}_s(k_{sx},k_{sy})e^{i\vec{k}_s\cdot\vec{r}_s}\,, \tag{S28}$$

where the subscript $s = i, r, t$ stands for incident, reflected and transmitted beams respectively. Moreover, we are working in the paraxial approximation, and so $k_{sz} = \sqrt{k_0^2 - (k_{sx}^2 + k_{sy}^2)} \approx k_0(1 - (k_{sx}^2 + k_{sy}^2)/2k_0^2)$, with $k_0$ being the central wave vector. Also, $\vec{\mathcal{E}}_s(k_{sx}, k_{sy})$ is the angular spectrum,

$$\vec{\mathcal{E}}_s = (\cos\beta\hat{x_s} + \sin\beta e^{i\varphi}\hat{y_s})\frac{w_0}{\sqrt{2\pi}}\exp\left(-w_0^2\frac{k_{sx}^2+k_{sy}^2}{4}\right)\,, \tag{S29}$$

where $\beta$ is the polarization angle, $\varphi$ is the phase difference between the $p$ polarized and the $s$ polarized field components (which are along $\hat{x}$ and $\hat{y}$ respectively in our case), $w_0$ is the beam waist, while $k_{sx}$ and $k_{sy}$ are the components of the wavevector $\vec{k}$ along $\hat{x}$ and $\hat{y}$ directions respectively. For convenience and ease of notation, we use the Dirac notation:

$$\vec{E}_s \equiv |E_s\rangle \quad\text{and}\quad \vec{\mathcal{E}}_s \equiv |\mathcal{E}_s\rangle\ . \tag{S30}$$

Upon reflection from the surface, the incident plane polarized beam splits into a right circularly polarized ($|E_r^+\rangle$) and a left circularly polarized ($|E_r^-\rangle$) beams spatially separated along the $\hat{y}$ direction [S2, S6]. This shift is given by [S7–S9],

$$\Delta_{y\pm} = \frac{\langle E_r^\pm|y|E_r^\pm\rangle}{\langle E_r^\pm|E_r^\pm\rangle} = \frac{\int dx_r dy_r\ y\ \vec{E}^*_{r\pm}\vec{E}_{r\pm}}{\int dx_r dy_r\ \vec{E}^*_{r\pm}\vec{E}_{r\pm}}\,. \tag{S31}$$

Equivalently,

$$\Delta_{y\pm} = \frac{\langle\mathcal{E}_r^\pm|i\frac{\partial}{\partial k_{ry}}|\mathcal{E}_r^\pm\rangle}{\langle\mathcal{E}_r^\pm|\mathcal{E}_r^\pm\rangle} = \frac{\int dk_{rx}dk_{ry}\ \vec{\mathcal{E}}^*_{r\pm}\ i\frac{\partial}{\partial k_{ry}}\vec{\mathcal{E}}_{r\pm}}{\int dk_{rx}dk_{ry}\ \vec{\mathcal{E}}^*_{r\pm}\vec{\mathcal{E}}_{r\pm}}\,, \tag{S32}$$

where

$$|\mathcal{E}_r^\pm\rangle = \widetilde{\mathcal{E}}_r^\pm|\pm\rangle|\mathcal{E}_i\rangle, \tag{S33}$$

with $\widetilde{\mathcal{E}}_r^\pm$ being the amplitude of the reflected beam obtained using the transfer matrix. In particular, for the $H$ polarization ($\beta = 0$ i.e. incident beam in the $p$ polarization), this amplitude in terms of the Fresnel coefficient is given by

$$\widetilde{\mathcal{E}}_r^\pm = \frac{1}{\sqrt{2}}\left[r_{pp} \mp ir_{sp} \pm i\frac{k_{ry}}{k_0}\cot\theta_i(r_{pp}+r_{ss})\right]\,. \tag{S34}$$

Note that $k_{ix} = -k_{rx}\,, k_{iy} = k_{ry}$. In the following, we explicitly quote the results only for the $H$ polarization (i.e. incident beam in the $p$ polarization); results for the incident $V$ polarization (i.e. incident beam in the $s$ polarization) are obtained analogously. We can now express the photonic-spin Hall shift for the RCP (+) and the LCP (−) along the $\hat{y}$ direction in terms of the Fresnel coefficients as [S8, S9],

$$\Delta^H_\pm = \mp\frac{\cot(\theta_i)}{k_0}Re\left[\frac{r_{pp}+r_{ss}}{r_{pp}\mp ir_{sp}}\right]\,. \tag{S35}$$

Note that here we have neglected the polarization-independent contribution to the shift since it is the same for both the RCP and LCP reflected beams. Using the optical conductivity dependence of the Fresnel coefficients we can write,

$$\Delta^H_\pm = \mp\frac{\cot(\theta_i)}{k_0}\frac{\mathcal{N}_\pm(\theta_i)}{\mathcal{D}_\pm(\theta_i)}\,, \tag{S36}$$

where

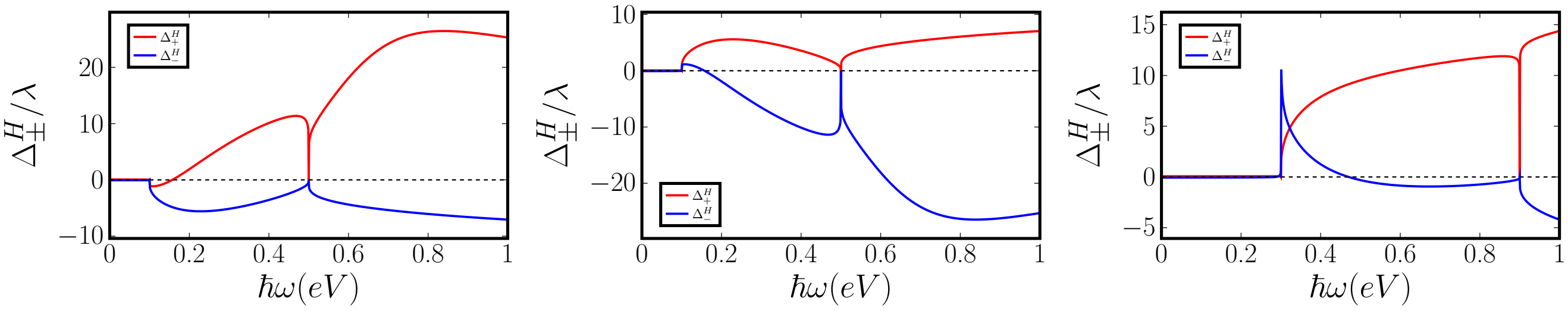


(a) (b) (c)

FIG. S1. Photonic spin-Hall shift for the RCP and the LCP reflected light in the case of Haldane model. (a) $M = 0.1\,\text{eV}$, $\phi = \pi/2$ and $C = 1$. (b) $M = 0.1\,\text{eV}$, $\phi = -\pi/2$ and $C = -1$. (c) $M = 0.3\,\text{eV}$, $\phi = \pi/2$ and $C = 0$. In all the plots, we have set $t' = 0.05/\sqrt{3}$ eV, $\mu = (\mu_a + \mu_b)/2 = 0$ eV. As discussed in the text, there is no clear sign structure in $\Delta_\pm$ to distinguish between the trivial and the topological phases.

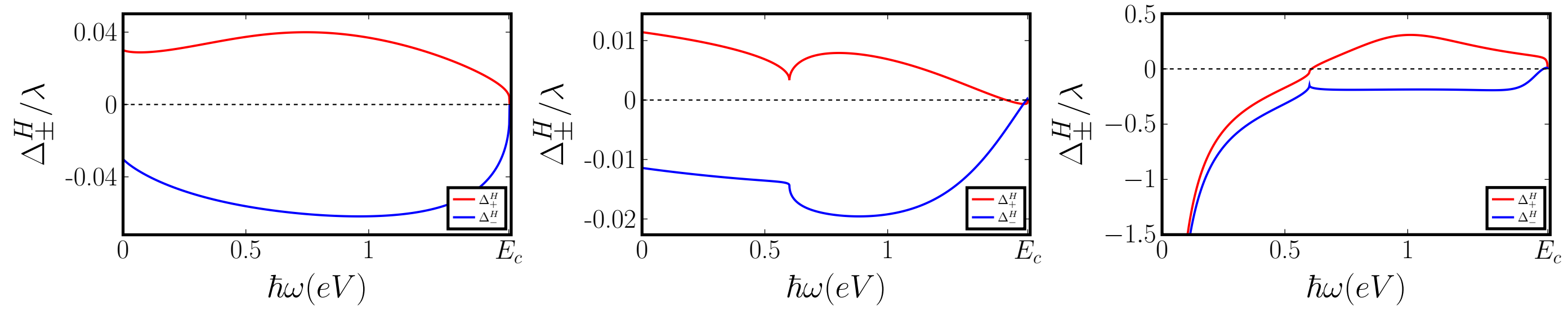


(a) (b) (c)

FIG. S2. RCP and LCP shift with the incident plane being the $x-y$ plane. (a) WSM phase. (b) QAH Phase. (c) NI phase. Here, $|\alpha| = 0.3eV$, $|\beta| = 0.25a^2eV\AA^2$ and $a = 2\AA$.

$$\mathcal{N}_\pm(\theta_i) = \big(f_0(\theta_i) + f_1(\theta_i)\sigma_{xx}^R\big)\left[f_2(\theta_i) + f_3(\theta_i)\sigma_{xx}^R \mp f_4(\theta_i)\sigma_{xy}^I + f_5(\theta_i)\Big([\sigma_{xx}^R]^2 - [\sigma_{xx}^I]^2 + [\sigma_{xy}^R]^2 - [\sigma_{xy}^I]^2\Big)\right]$$
$$+ f_1(\theta_i)\sigma_{xx}^I\left[f_3(\theta_i)\sigma_{xx}^I \pm f_4(\theta_i)\sigma_{xy}^R + 2f_5(\theta_i)\Big(\sigma_{xx}^R\sigma_{xx}^I + \sigma_{xy}^R\sigma_{xy}^I\Big)\right], \tag{S37}$$

$$\mathcal{D}_\pm(\theta_i) = \left[f_2(\theta_i) + f_3(\theta_i)\sigma_{xx}^R \mp f_4(\theta_i)\sigma_{xy}^I + f_5(\theta_i)\Big([\sigma_{xx}^R]^2 - [\sigma_{xx}^I]^2 + [\sigma_{xy}^R]^2 - [\sigma_{xy}^I]^2\Big)\right]^2$$
$$+ \left[f_3(\theta_i)\sigma_{xx}^I \pm f_4(\theta_i)\sigma_{xy}^R + 2f_5(\theta_i)\Big(\sigma_{xx}^R\sigma_{xx}^I + \sigma_{xy}^R\sigma_{xy}^I\Big)\right]^2 > 0, \tag{S38}$$

with

$$f_0(\theta_i) = -\frac{2\epsilon_0 c\sin^2\theta_i}{n_2}(n_2^2 - 1) \tag{S39}$$

$$f_1(\theta_i) = -2\sin^2\theta_i\sqrt{1 - \left(\frac{\sin\theta_i}{n_2}\right)^2} \tag{S40}$$

$$f_2(\theta_i) = \epsilon_0 c\frac{n_2^2 - 1}{n_2}(\cos\theta_i\sqrt{n_2^2 - \sin^2\theta_i} - \sin^2\theta_i) \tag{S41}$$

$$f_3(\theta_i) = \frac{\cos\theta_i}{n_2}(2n_2^2 - \sin^2\theta_i) - \sin^2\theta_i\sqrt{1 - \left(\frac{\sin\theta_i}{n_2}\right)^2} \tag{S42}$$

$$f_4(\theta_i) = 2\cos\theta_i\sqrt{\left(1 - \frac{\sin^2\theta_i}{n_2^2}\right)} \tag{S43}$$

$$f_5(\theta_i) = \mu_0 c\cos\theta_i\sqrt{1-\left(\frac{\sin\theta_i}{n_2}\right)^2}\,. \tag{S44}$$

At Brewster's angle,

$$f_0(\theta_B) = -\frac{2n_2\epsilon_0 c(n_2^2-1)}{n_2^2+1} \tag{S45}$$

$$f_1(\theta_B) = -\frac{2n_2^3}{(n_2^2+1)^{\frac{3}{2}}} \tag{S46}$$

$$f_2(\theta_B) = 0 \tag{S47}$$

$$f_3(\theta_B) = \frac{n_2}{\sqrt{n_2^2+1}} \tag{S48}$$

$$f_4(\theta_B) = \frac{2n_2}{n_2^2+1} \tag{S49}$$

$$f_5(\theta_B) = \frac{\mu_0 c n_2}{n_2^2+1}\,. \tag{S50}$$

It is clear from the above expressions that the individual polarization shifts are non-trivial functions of both the longitudinal and the Hall conductivities. This makes it difficult to single out the role of the Hall conductivity, which is an essential quantity characterizing the band topology when time-reversal is broken. Therefore, we consider the centroid shift, defined as

$$\Delta_c = \frac{\langle\mathcal{E}_r^H|i\frac{\partial}{\partial k_{ry}}|\mathcal{E}_r^H\rangle}{\langle\mathcal{E}_r^H|\mathcal{E}_r^H\rangle}\,, \tag{S51}$$

which is the intensity weighted average of the individual shifts corresponding to the right and the left circularly polarized reflected light,

$$\Delta_c = \frac{I_+\Delta_+^H + I_-\Delta_-^H}{I_+ + I_-}\,, \tag{S52}$$

where $I_\pm$ are the respective intensities of the RCP/LCP light. In terms of the Fresnel coefficients,

$$\Delta_c = -\frac{\cot(\theta_i)}{k_0} Re\left[\frac{i(r_{pp}+r_{ss})r_{sp}^*}{|r_{pp}|^2+|r_{sp}|^2}\right]. \tag{S53}$$

Such an observable is possible to measure in experiments, see for instance Ref. [S10].

Again, inserting the Fresnel coefficient expressions in terms of conductivities, we obtain the central expression of this work,

$$\Delta_c = -\frac{\cot(\theta_i)}{k_0}\frac{N_c}{D_c}\,, \tag{S54}$$

where

$$N_c = g_1\left[g_2(\sigma_{xx}^R\sigma_{xy}^I - \sigma_{xx}^I\sigma_{xy}^R) + g_3\sigma_{xy}^I\right], \tag{S55}$$

$$\begin{aligned} D_c = {} & g_1^2\left([\sigma_{xy}^R]^2 + [\sigma_{xy}^I]^2\right) \\ & + \left[g_4 + g_5\sigma_{xx}^R + g_6\left([\sigma_{xx}^R]^2 - [\sigma_{xx}^I]^2 + [\sigma_{xy}^R]^2 - [\sigma_{xy}^I]^2\right)\right]^2 \\ & + \left[g_5\sigma_{xx}^I + 2g_6(\sigma_{xx}^I\sigma_{xx}^R + \sigma_{xy}^I\sigma_{xy}^R)\right]^2 > 0\,, \end{aligned} \tag{S56}$$

with,

$$g_1 = 2\cos\theta_i\sqrt{1-\left(\frac{\sin\theta_i}{n_2}\right)^2} \tag{S57}$$

$$g_2 = 2\sin^2\theta_i\sqrt{1-\left(\frac{\sin\theta_i}{n_2}\right)^2} \tag{S58}$$

$$g_3 = \frac{2\epsilon_0 c\sin^2\theta_i}{n_2}(n_2^2-1) \tag{S59}$$

$$g_4 = \epsilon_0 c\frac{n_2^2-1}{n_2}(\cos\theta_i\sqrt{n_2^2-\sin^2\theta_i} - \sin^2\theta_i) \tag{S60}$$

$$g_5 = \frac{\cos\theta_i}{n_2}(2n_2^2-\sin^2\theta_i) - \sin^2\theta_i\sqrt{1-\left(\frac{\sin\theta_i}{n_2}\right)^2} \tag{S61}$$

$$g_6 = \mu_0 c\cos\theta_i\sqrt{1-\left(\frac{\sin\theta_i}{n_2}\right)^2}\,. \tag{S62}$$

At Brewster's angle the above constants are

$$g_1 = \frac{2n_2}{n_2^2+1}\,, \tag{S63}$$

$$g_2 = \frac{2n_2^3}{(n_2^2+1)^{\frac{3}{2}}}\,, \tag{S64}$$

$$g_3 = \frac{2\epsilon_0 c n_2}{n_2^2+1}(n_2^2-1)\,, \tag{S65}$$

$$g_4 = 0\,, \tag{S66}$$

$$g_5 = \frac{n_2}{\sqrt{n_2^2+1}}\,, \tag{S67}$$

$$g_6 = \frac{\mu_0 c n_2}{n_2^2+1}\,. \tag{S68}$$

Eq. (S55) is the central result, which shows the direct dependence of the centroid shift on the Hall conductivity. This leads to specific sign structure as a function of frequency, as discussed in the main text. The centroid shift for specific cases is shown in Figs. 2 and 3 in the main text. In Figs. (S3) and (S4) we show the corresponding variation with the incident angle of light beam. Throughout this work, we consider the substrate to be glass, $n_2 = 1.5$. However, our general results and conclusions do not depend on the choice of the substrate.

## III. CONDUCTIVITY

The optical conductivity for $2D$ system is calculated within the linear response theory using the Kubo-Greenwood formula [S11]:

$$\sigma_{ij}(\omega) = \frac{e^2\hbar}{i}\int\frac{d^2\vec{k}}{(2\pi)^2}\sum_{b,b'}\lim_{\vec{k}'\to\vec{k}}\left[\frac{f(E_b(\vec{k})) - f(E_{b'}(\vec{k}'))}{E_b(\vec{k}) - E_{b'}(\vec{k}')}\right.$$

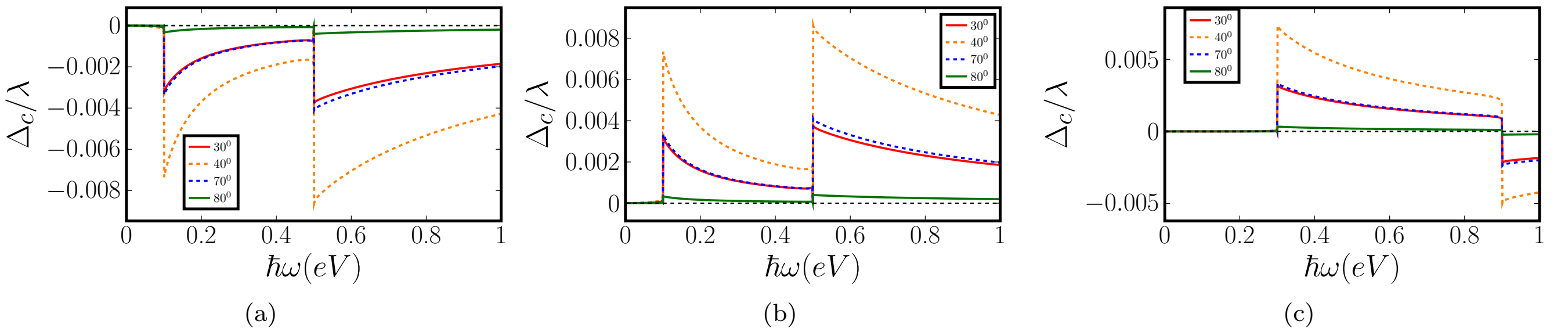


FIG. S3. Frequency dependence of the centroid shift at different incident angles of the light beam. Parameters same as in Fig. (S1).

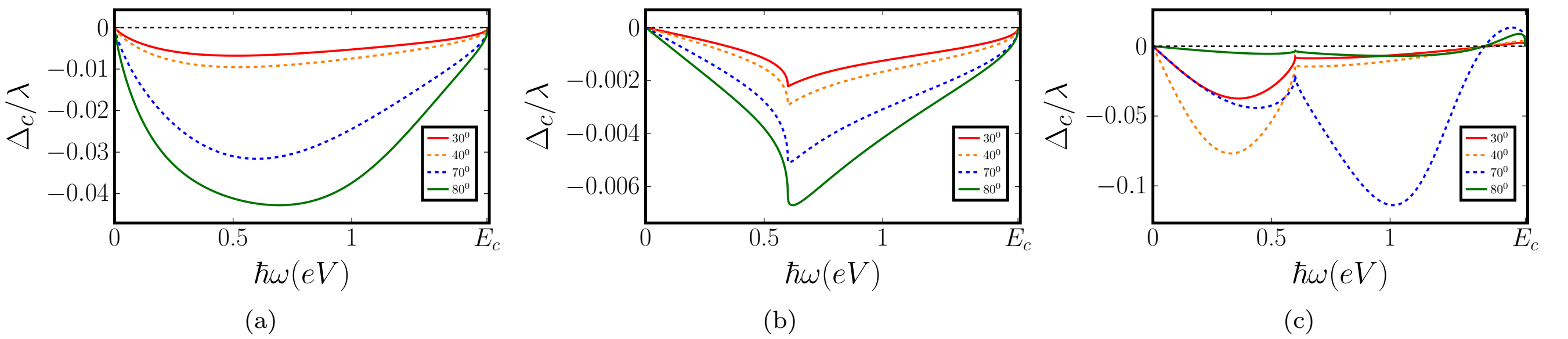


FIG. S4. Centroid shift obtained upon reflection of incident light from the $x-y$ plan in different $3d$ phases, plotted for different incident angles. (a) WSM phase. (b) QAH phase. (c) NI phase. Parameters same as in Fig. (S2).

$$\times \frac{\langle\psi_b(\vec{k})|\hat{v_i}|\psi_{b'}(\vec{k}')\rangle\langle\psi_{b'}(\vec{k}')|\hat{v_j}|\psi_b(\vec{k})\rangle}{E_b(\vec{k})-E_{b'}(\vec{k}')+\hbar\omega}\Bigg], \qquad \text{(S69)}$$

where $i,j=x,y$ and $b,b'$ are the energy band indices (eg. conduction band and valence band) with $E_b, E_b'$ the corresponding energy eigenvalues. Also, $f(\varepsilon)$ is the Fermi-Dirac distribution function. Specifically, the optical conductivity has two contributions,

$$\sigma_{ij}(\omega)=\sigma_{ij}^{intra}(\omega)+\sigma_{ij}^{inter}(\omega)\,, \qquad \text{(S70)}$$

where $\sigma_{ij}^{intra}$ ($\sigma_{ij}^{inter}$) is the intraband (interband) contribution to the conductivity.

### A. Chern insulator

Below, we briefly recall the results for the Haldane model. We consider the low-energy Hamiltonian (Eq. (6) in the main text) describing the physics around the two valleys, $K$ and $K'$ points. Neglecting the intervalley scattering processes and at $T=0$ K, we obtain [S12],

$$\sigma_{xx}^{intra}(\omega)=\frac{e^2}{4\hbar}\frac{(\mu+\lambda)^2-\Delta_\tau^2}{(\mu+\lambda)}\left[\delta(\hbar\omega)+\frac{i}{\pi\hbar\omega}\right]\Theta(\mu-E_c^\tau)$$

$$=\sigma_{yy}^{intra}(\omega)\,, \qquad \text{(S71)}$$

$$\sigma_{xy}^{intra}(\omega)=\sigma_{yx}^{intra}(\omega)=0\,, \qquad \text{(S72)}$$

$$\sigma_{xx}^{inter}(\omega)=\frac{e^2}{16\hbar}\left[1+\frac{4\Delta_\tau^2}{(\hbar\omega)^2}\right]\Bigg[\Theta(\hbar\omega-2(\Lambda_\tau+\lambda))$$

$$+\frac{i}{\pi}\left(\ln\left|\frac{2(\Lambda_\tau+\lambda)-\hbar\omega}{2(\Lambda_\tau+\lambda)+\hbar\omega}\right|+\frac{1}{\hbar\omega}\frac{4\Delta_\tau^2}{\Lambda_\tau+\lambda}\right)\Bigg]$$

$$=\sigma_{yy}^{inter}(\omega)\,, \qquad \text{(S73)}$$

$$\sigma_{xy}^{inter}(\omega)=\frac{e^2\tau\Delta_\tau}{4\pi\hbar^2\omega}$$

$$\times\left[\ln\left|\frac{2(\Lambda_\tau+\lambda)-\hbar\omega}{2(\Lambda_\tau+\lambda)+\hbar\omega}\right|-i\pi\Theta(\hbar\omega-2(\Lambda_\tau+\lambda))\right]$$

$$=-\sigma_{yx}^{inter}(\omega)\,, \qquad \text{(S74)}$$

where $\tau=\pm1$ is the valley index corresponding to the $K$ and $K'$ respectively, and $\Theta(x)$ is the Heaviside function. In the above expressions, $\mu$ is the Fermi energy (throughout this work $\mu=0$), $\Lambda_\tau = max[\mu, E_c^\tau]$ and $E_c^\tau=-\lambda+|\Delta_\tau|$ is the energy of the bottom of the conduction band. Our results are consistent with Ref. [S13] in the appropriate limit. In this work, the Fermi level is such that the lower band is always fully filled. So the intraband contribution to the conductivity is always zero. We use the notation $\sigma_{ij}(\omega)=\sigma_{ij}^R+i\sigma_{ij}^I$ to denote the real and imaginary parts of the optical conductivity. In Figs. S6, we have plotted the optical conductivities for the Haldane model in different regimes. In the dc limit, $\sigma_{xy}=C\sigma_0$, where $C$ is the Chern number and $\sigma_0=e^2/h$.

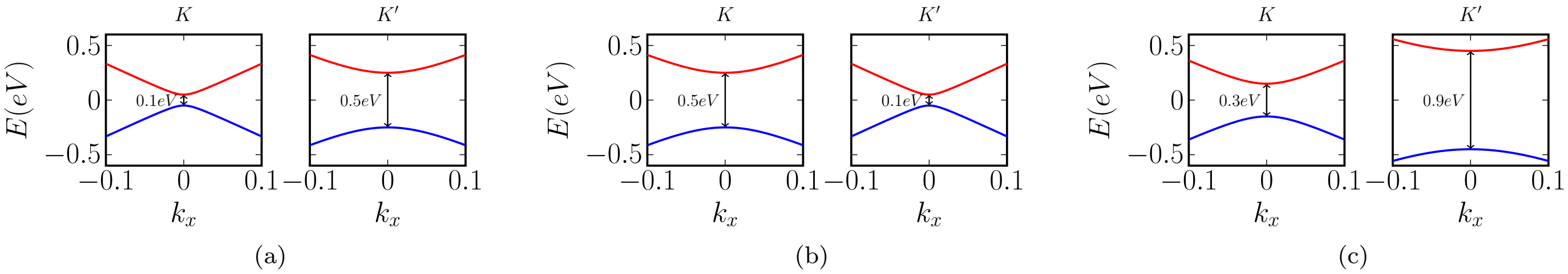


FIG. S5. Energy bands near the two valleys in the case of Haldane model. Parameters same as in Fig. (S1). Additionally, here $a = 1.42\mathring{A}\,,\ t = 2.7eV$.

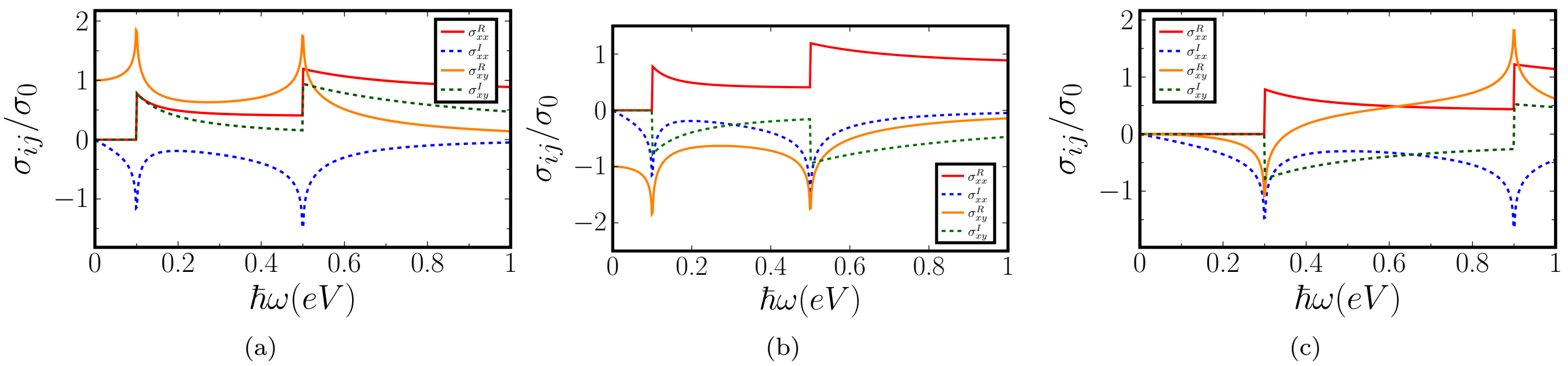


FIG. S6. Optical conductivities for the Haldane model. Parameters same as in Fig. S1.

### B. Weyl semimetal

We consider the minimal model for Weyl semimetal with broken time-reversal symmetry, as shown in Eq. (7) in the main text. Using the Kubo formula, one can straightforwardly obtain the following optical conductivity expressions [S8, S9]:

$$\sigma_{xx}(\omega) = \frac{e^2 t\omega}{12\pi\hbar v_z} - i\frac{e^2 t\omega}{12\pi^2\hbar v_z}\ln\left|\frac{\omega^2-\omega_c^2}{\omega^2}\right| = \sigma_{yy}(\omega)\,, \tag{S75}$$

$$\begin{aligned}\sigma_{xy}(\omega) &= \frac{be^2t}{2\pi^2\hbar}\left(1+\frac{\omega^2}{12v_z^2(k_c^2-b^2)}\right) \\ &- i\left[\frac{e^2tb}{2\pi^3\hbar}\ln\left|\frac{\omega_c-\omega}{\omega_c+\omega}\right| \right.\\ &\left.+\frac{e^2\omega tb}{12\pi^3\hbar v_z^2(k_c^2-b^2)}\left(\omega_c+\frac{\omega}{2}\ln\left|\frac{\omega_c-\omega}{\omega_c+\omega}\right|\right)\right] \\ &= -\sigma_{yx}(\omega)\,,\end{aligned} \tag{S76}$$

where $k_c = \pi/a$ is the momentum cutoff along $z$-axis, $\omega_c = k_c v_z$, and $t$ is the thickness of the material. Note that $\sigma_{xz} = \sigma_{yz} = 0$ since the Weyl nodes are along the $k_z$ axis. In Fig. (S7)(a), we show the optical conductivities in the Weyl SM phase. Using these relations we can see that the numerator in the expression for the centroid shift, given in Eq. (4) of the main text (Eq. (S55) above), is always negative for any $\hbar\omega < E_c$. The angle dependence of the centroid shift in the WSM phase is shown in Fig. (S4)(a).

### C. Quantum Anomalous Hall (QAH) and Normal Insulator (NI) phases

The minimal model for the QAH and NI phases in $3d$ is written in Eq. (8) in the main text [S14]. The optical conductivity in both these phases is again obtained using the Kubo formula. It turns out that the longitudinal conductivity in both the $NI$ and $QAH$ phases is the same. The real part of the longitudinal conductivity is

$$\sigma^R_{xx} = \frac{e^2}{\hbar}\sqrt{\hbar\omega-2|\alpha|}\,\Theta(\hbar\omega-2|\alpha|)\left[\frac{A}{(\hbar\omega)^2}+\frac{B}{\hbar\omega}+C\right], \tag{S77}$$

where,

$$A = \frac{16|\alpha|^2 t}{120\sqrt{2}\pi\sqrt{|\beta|}}\,,\quad B = \frac{4|\alpha|t}{120\sqrt{2}\pi\sqrt{|\beta|}}\,,\quad C = \frac{9t}{120\sqrt{2}\pi\sqrt{|\beta|}}\,. \tag{S78}$$

with $\alpha$ and $\beta$ are parameters introduced in Eq. (8) in the main text. Similarly, the imaginary part of the longitudinal conductivity is given by,

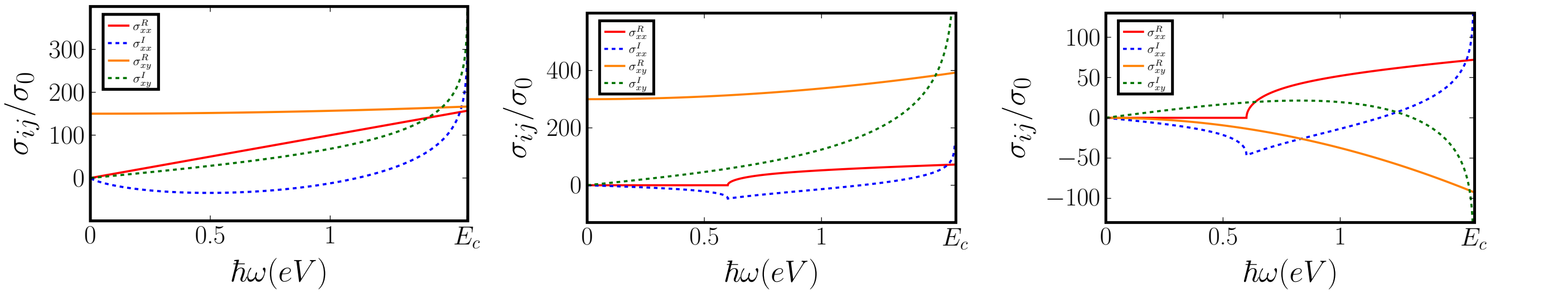


(a) (b) (c)

FIG. S7. Optical conductivity in the $xy$ plane in different phases in $3d$. (a) WSM phase. Here, $t = 60$ nm, $E_c = \hbar\omega_c = \pi/2$, with $a = 2$Å, and the Weyl node separation is $b = 0.5\ \pi/a$. (b) QAH phase with $t = 60nm$, Cutoff energy $E_c = \hbar\omega_c = \pi/2$. (c) NI phase with $t = 60nm$, Cutoff energy $E_c = \hbar\omega_c = \pi/2$.

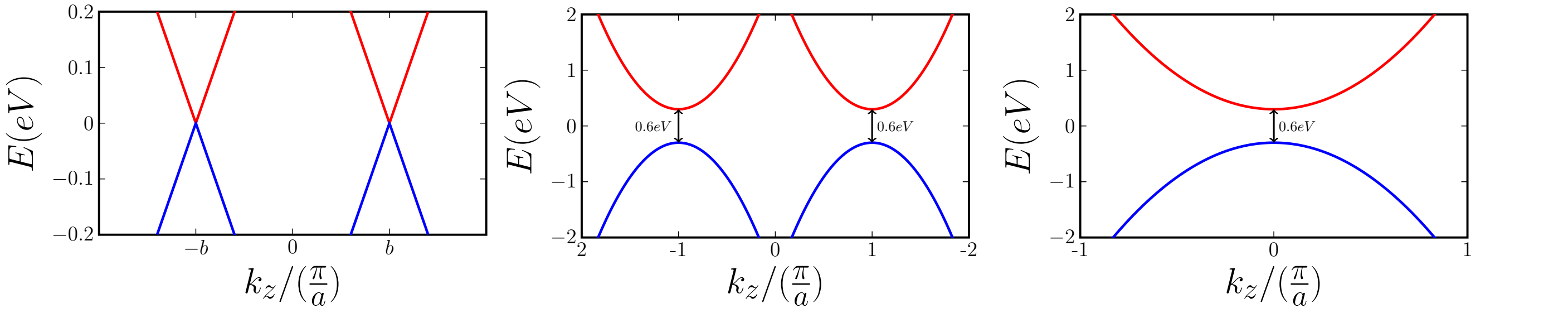


(a) (b) (c)

FIG. S8. Low-energy bands for different phases in $3d$. (a) WSM phase, $b = 0.5\pi/a$, with $a = 2$ Å. (b) QAH phase. (c) NI phase. Parameters same as in Fig. (S2).

$$\begin{aligned}\sigma_{xx}^{I} = \frac{e^2}{\hbar}\Bigg\{ &- \frac{2A}{\pi\hbar\omega}\frac{\sqrt{\hbar\omega_c - 2|\alpha|}}{\hbar\omega_c} + \left[\frac{4}{\pi}\left(\frac{A}{2\sqrt{2|\alpha|}\hbar\omega} - \frac{B\sqrt{2|\alpha|}}{\hbar\omega}\right)\right]\tan^{-1}\left(\sqrt{\frac{\hbar\omega_c - 2|\alpha|}{2|\alpha|}}\right)\\ &- \left[\frac{4}{\pi}\left(\frac{A}{\hbar\omega} + B + C(\hbar\omega)\right)\right]\left(\frac{\sqrt{\hbar\omega - 2|\alpha|}}{4\hbar\omega}\ln\left|\frac{\sqrt{\hbar\omega_c - 2|\alpha|} - \sqrt{\hbar\omega - 2|\alpha|}}{\sqrt{\hbar\omega_c - 2|\alpha|} + \sqrt{\hbar\omega - 2|\alpha|}}\right|\Theta[\hbar\omega - 2|\alpha|]\right.\\ &\left. - \frac{\sqrt{2|\alpha| - \hbar\omega}}{2\hbar\omega}\tan^{-1}\left(\sqrt{\frac{\hbar\omega_c - 2|\alpha|}{2|\alpha| - \hbar\omega}}\right)\Theta[2|\alpha| - \hbar\omega]\right)\\ &- \left[\frac{4}{\pi}\left(\frac{A}{\hbar\omega} - B + C(\hbar\omega)\right)\right]\frac{\sqrt{\hbar\omega + 2|\alpha|}}{2\hbar\omega}\tan^{-1}\left(\sqrt{\frac{\hbar\omega_c - 2|\alpha|}{\hbar\omega + 2|\alpha|}}\right)\Bigg\}\end{aligned} \tag{S79}$$

Next, we quote the results for the Hall conductivity in the QAH phase,

$$\sigma_{xy}^{R} = \frac{e^2}{\hbar}\left[\frac{k_c t}{2\pi^2} - A_{xy}^{NI}(\hbar\omega)^2\right], \tag{S80}$$

$$\begin{aligned}\sigma_{xy}^{I} = -\frac{e^2}{\hbar}\Bigg\{&\frac{k_c t}{2\pi^3}\ln\left|\frac{\omega_c - \omega}{\omega_c + \omega}\right|\\ &- A_{xy}^{NI}\frac{2\hbar\omega}{\pi}\left[\hbar\omega_c + \frac{\hbar\omega}{2}\ln\left|\frac{\omega_c - \omega}{\omega_c + \omega}\right|\right]\Bigg\}.\end{aligned} \tag{S81}$$

Similarly, in the NI phase,

$$\sigma_{xy}^{R} = \frac{e^2}{\hbar}A_{xy}^{NI}(\hbar\omega)^2, \tag{S82}$$

$$\sigma_{xy}^{I} = -\frac{e^2}{\hbar}A_{xy}^{NI}\frac{2\hbar\omega}{\pi}\left[\hbar\omega_c + \frac{\hbar\omega}{2}\ln\left|\frac{\omega_c - \omega}{\omega_c + \omega}\right|\right]. \tag{S83}$$

In the above expressions,

$$A_{xy}^{NI} = -\frac{k_c t}{96\pi^2\alpha(\alpha+\beta k_c^2)} - \frac{\tan^{-1}\left(\sqrt{\frac{\beta k_c^2}{\alpha}}\right)t}{96\pi^2\sqrt{\alpha^3\beta}}, \quad \text{(S84)}$$

with $k_c = \pi/a$ being the momentum cut off along the $z$ axis. Using the above expressions, we can again see that in the QAH phase the numerator in the centroid shift expression is always negative for all $\hbar\omega < E_c$. On the other hand, in the NI phase, the numerator necessarily changes sign for $\hbar\omega < E_c$. The precise determination of the frequency where the sign changes is however difficult.



\end{document}